\documentclass[final,5p,times,twocolumn]{elsarticle}

 \usepackage{graphicx}

\usepackage{amssymb}

 \biboptions{sort&compress}

\begin{document}

\begin{frontmatter}



\title{Concerning the Time Dependence of the Decay Rate of $^{137}$Cs}


\author[NE,phy]{J. H. Jenkins\corref{cor}}
\address[NE]{School of Nuclear Engineering, Purdue University, West Lafayette, IN 47907, USA}
\cortext[cor]{Corresponding author}
\ead{jere@purdue.edu}
\author[phy]{E. Fischbach}
\address[phy]{Department of Physics, Purdue University, West Lafayette, IN 47907, USA}
\author[412]{D. Javorsek, II}
\address[412]{412th Test Wing, Edwards AFB, CA 93524, USA}
\author[USAFA]{Robert H. Lee}
\address[USAFA]{Physics Department; U.S. Air Force Academy; 2354 Fairchild Dr., USAFA, CO 80840 USA}
\author[stan]{P. A. Sturrock}
\address[stan]{Center for Space Science and Astrophysics, Stanford University, Stanford, CA 94305, USA}

\begin{abstract}
The decay rates of 8 nuclides ($^{85}$Kr, $^{90}$Sr, $^{108}$Ag, $^{133}$Ba, $^{137}$Cs, $^{152}$Eu, $^{154}$Eu, and $^{226}$Ra) were monitored by the standards group at the Physikalisch-Technische Bundesanstalt (PTB), Braunschweig, Germany, over the time frame June 1999 to November 2008. We find that the PTB measurements of the decay rate of $^{137}$Cs show no evidence of an annual oscillation, in agreement with the recent report by Bellotti et al. However, power spectrum analysis of PTB measurements of a $^{133}$Ba standard, measured in the same detector system, does show such evidence. This result is consistent with our finding that different nuclides have different sensitivities to whatever external influences are responsible for the observed periodic variations.

\end{abstract}

\begin{keyword}

Radioactivity \sep Beta Decay \sep Sun \sep Neutrinos \sep Gran Sasso
\end{keyword}

\end{frontmatter}



\section{Introduction}
\label{sec:Intro}

\citet{bel12} have recently reported the result of measurements of the activity of a $^{137}$Cs source (T$_{1/2}$=30.08 y, 100\% $\beta^{-}$ \citep{Cs137}), as determined by an experiment installed deep underground in the Laboratori Nazionali del Gran Sasso (LNGS). They report that ``no signal with amplitude larger than 9.6$\times{}10^{-5}$ at 95\% C.L. has been detected,'' concluding that this result is ``in clear contradiction with previous experimental results and their interpretation as indication of a novel field (or particle) from the Sun.'' In reviewing the case for variability, Bellotti et al. refer to articles by \citet{jen09b}, \citet{fis11}, \citet{par10a,par10b}, and \citet{jav10}. However, none of these articles cites decay rates for $^{137}$Cs. 

Measured decay data for $^{137}$Cs have previously been examined for similar periodicities, and none have been found. A small source of $^{137}$Cs is onboard the MESSENGER spacecraft, and decay data have been collected periodically from an onboard high purity germanium detector. These data, from just prior to launch on Earth to just after orbital insertion at Mercury, are consistent with no modulation of $^{137}$Cs, as reported by \citet{fis12}. \citet{ell90} also reported no annual or other variations in measured $^{137}$Cs decay data. What is striking about Ellis'
 result, however, is that there was an annual variation in the measurements of $^{56}$Mn decay data, which were taken on the same detector system (over the same time period) that was used to measure the $^{137}$Cs calibration standards, for which no annual oscillatory behavior was observed. We shall discuss the Ellis results in greater detail in Section \ref{sec:Disc}. There is one group  that has reported periodicities shorter than a year in $^{137}$Cs, \citep{bau00,bau01}, but none of those experiments had a long enough duration to conclusively observe an annual period.

The question of periodic or other non-random behaviors in nuclear decay rates has long been of interest to the scientific community \citep{eme72,hah76,dos77}, and the results presented in Refs. \citep{jen09a,jen09b,jav10,stu10a,stu10b}, as well as others, have generated renewed interest in this topic. In an effort to further explore the possible existence of periodicities in nuclear decays, an examination of historical data collected during extended studies of half-lives of long-lived radionuclides and of detector stability has been carried out at the Physikalisch-Technische Bundesanstalt (PTB). Included in this program was an analysis of $^{137}$Cs data, as reported by \citet{sch10}. The goal of the present article is to examine the results of these PTB measurements for comparison with the result of the Bellotti experiment.

\section{Analysis of PTB measurements}
\label{sec:Analysis}
\citet{sch10} reports extended half-life measurements of 8 nuclides ($^{85}$Kr, $^{90}$Sr, $^{108}$Ag, $^{133}$Ba, $^{137}$Cs, $^{152}$Eu, $^{154}$Eu, and $^{226}$Ra). The $^{137}$Cs data were collected from late 1998 to November 2008. All of the measurements were made with a 4$\pi$ ionization chamber (IG12/A20, Centronic 20th Century Electronics, Ltd.). In principle, these measurements could be affected by influences on the particular radionuclide under study, on the detector, or on the measuring electronics. As is clear from Figures 2, 3, 4, and 5 of \citet{sch10}, the record of measurements (residuals of a half-life fit) superficially resembles a scatter diagram. It follows that periodicities in the decays of any of these nuclides will be revealed only by some form of power-spectrum analysis. The purpose of the present article is to carry out such an analysis for $^{137}$Cs (residuals of a half-life fit) and also (for reasons that will become clear) for $^{133}$Ba and $^{226}$Ra.

Since oscillations--when they occur--are typically intermittent rather than steady, it is more illuminating to examine time-frequency displays (``spectrograms'') than simple power spectra \citep{stu08}. To form spectrograms, we first prepare the data by means of the RONO (Rank-Order NOrmalization) operation \citep{stu11a} that maps the measurements onto a normal distribution, as is appropriate for power-spectrum analyses such as the Lomb-Scargle procedure \citep{lom76,sca82} or a likelihood procedure \citep{stu06}. We then carry out a sequence of likelihood power-spectrum analyses of sections of the data. For present purposes, we have found it convenient to adopt sections of 500 measurements. The power, $S$, is then displayed by a color code in a time-frequency diagram. (In power-spectrum analysis, the probability of finding a power of $S$ or greater at a given frequency arising from  normally distributed random noise, the null hypothesis, is given by $e^{-S}$ \cite{sca82}).

\begin{figure}[h]
\includegraphics[width=\columnwidth]{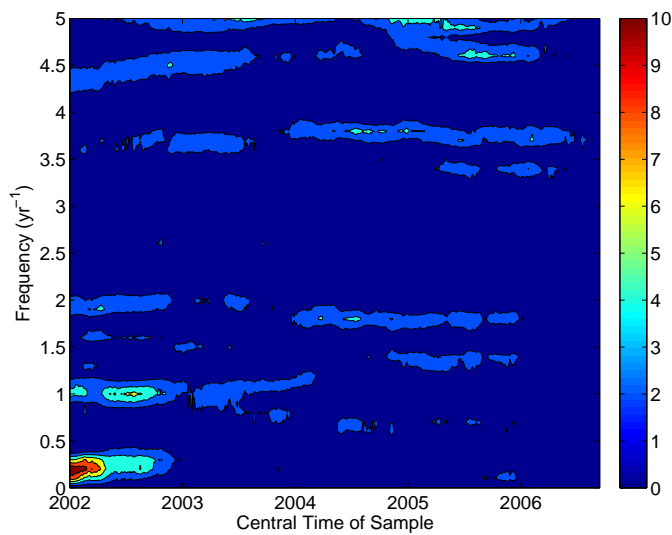}
\caption{Time-frequency display (spectrogram) of measurements of the decay-rate of $^{137}$Cs made at PTB over the time interval June 1999 to November 2008. There is only a slight suggestion of an annual oscillation from 2002 to 2003. The power, $S$, is represented by the color bar. \label{fig:fig1}}
\end{figure}

\begin{figure}[h]
\includegraphics[width=\columnwidth]{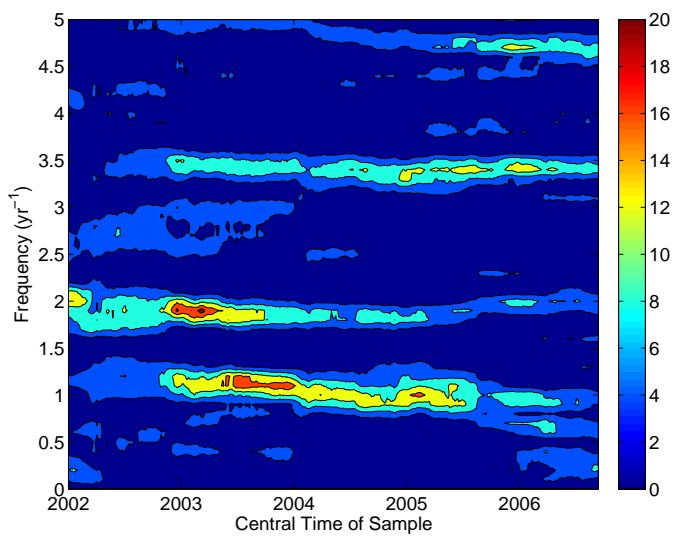}
\caption{Spectrogram of measurements of the decay-rate of $^{133}$Ba made at PTB over the time interval June 1999 to November 2008. There is evidence of an annual oscillation from 2003 to 2005. There is also evidence of the first harmonic of this oscillation. The power, $S$, is represented by the color bar. \label{fig:fig2}}
\end{figure}

The spectrogram formed in this way from the PTB $^{137}$Cs data is shown in Figure \ref{fig:fig1}. We see that there is only slight evidence of an annual oscillation (frequency 1 year$^{-1}$) between 2002 and 2004. The feature near 0.2 year$^{-1}$ may be related to the finite duration of the dataset. In contrast, we show in Figure \ref{fig:fig2} the spectrogram formed from the $^{133}$Ba (T$_{1/2}$=10.551 y, 100\% K-capture \citep{Ba133}) measurements taken on the same detector system. This spectrogram exhibits a strong annual oscillation from 2003 to 2005. We also see evidence of an oscillation with frequency close to 2 year$^{-1}$.  This could be a harmonic of the annual oscillation, but it is more likely to be a Rieger oscillation (an r-mode oscillation with spherical harmonic indicies $l=3,m=1$), which is prominent in power spectra formed from Brookhaven National Laboratory (BNL) and PTB data \citep{stu11a}.

\begin{figure}[h]
\includegraphics[width=\columnwidth]{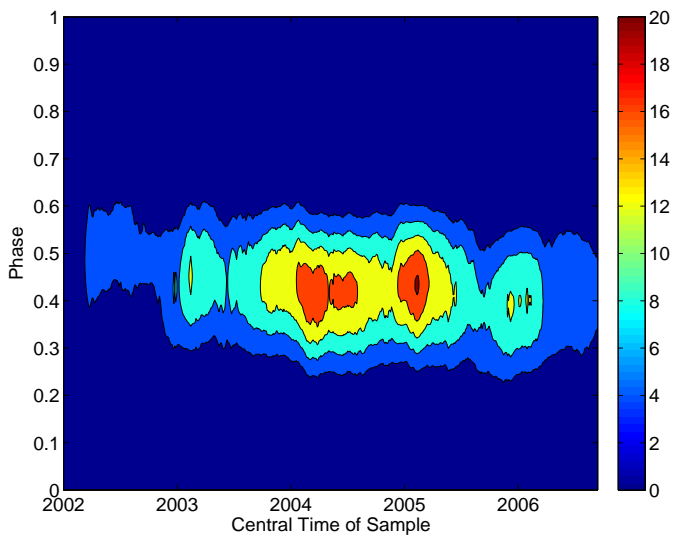}
\caption{Time-phase display (phasegram) of measurements of the decay-rate of $^{133}$Ba made at PTB over the time interval June 1999 to November 2008. The phase of the annual oscillations is approximately 0.43, corresponding to a peak in the modulation on or about June 6. The power, $S$, is represented by the color bar. \label{fig:fig3}}
\end{figure}

We have also analyzed the PTB measurements in terms of ``phasegrams,'' which are analogous to spectrograms, displaying the power as a function of time and phase for an assumed annual oscillation. The plot derived from $^{133}$Ba data is shown in Figure \ref{fig:fig3}. As we expect from Figure \ref{fig:fig2}, the power is found mainly over the time interval 2003 to 2005. The power is centered on a phase of approximately 0.43, corresponding to a date on or about 6 June.

\begin{figure}[h]
\includegraphics[width=\columnwidth]{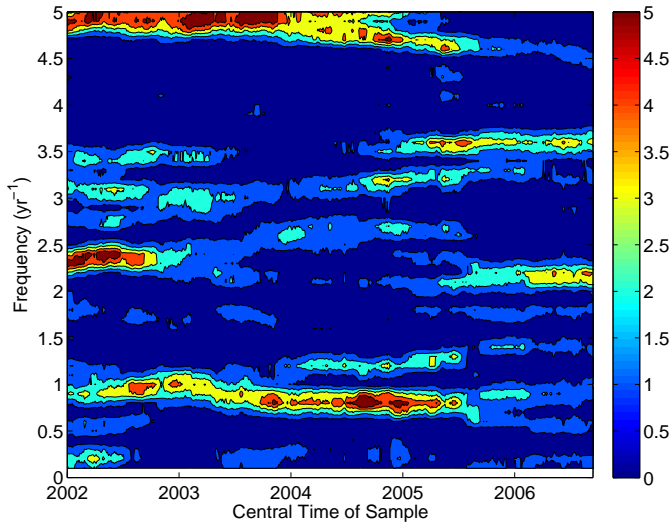}
\caption{Spectrogram of measurements of the decay-rate of $^{226}$Ra made at PTB over the time interval June 1999 to November 2008. There is evidence of an annual oscillation from 2002 to 2005. The power, $S$, is represented by the color bar. \label{fig:fig4}}
\end{figure}

\begin{figure}[h]
\includegraphics[width=\columnwidth]{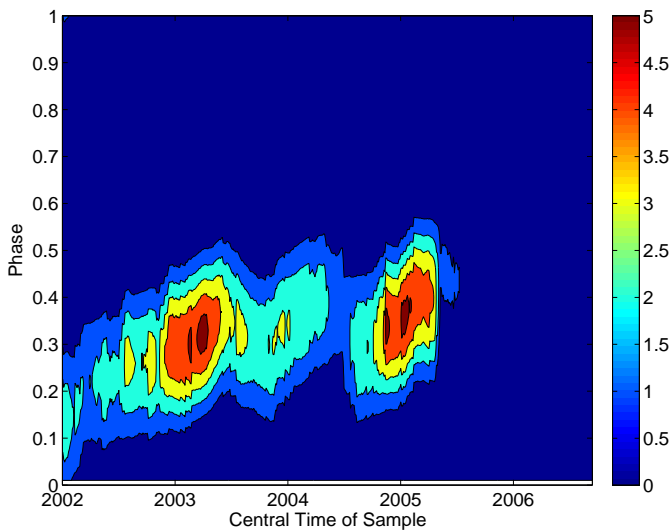}
\caption{Time-phase display (phasegram) of measurements of the decay-rate of $^{226}$Ra made at PTB over the time interval June 1999 to November 2008. The phase of the annual oscillation is approximately 0.36, corresponding to a peak in the modulation on or about 12 May. The power, $S$, is represented by the color bar. \label{fig:fig5}}
\end{figure}

Comparable plots for $^{226}$Ra (T$_{1/2}$=1600 y, 100\% $\alpha$-decay \citep{Ra226}) measurements are shown in Figures \ref{fig:fig4} and \ref{fig:fig5}. We see from these figures that the $^{226}$Ra results are similar to those for $^{133}$Ba, but the power levels are not as strong. We note that while $^{226}$Ra is 100\% $\alpha$-decay, it is in equilibrium with most of its daughters, several of which are $\beta^{-}$-decays. These $\beta^{-}$-decaying daughters contribute a significant portion of the photons emanating from the sealed source \citep{chi07}. Therefore, we cannot discern which isotope or isotopes would be the source of the observed fluctuations and note that there could be more than one.

\section{Discussion}
\label{sec:Disc}
We next review in more detail previous reports concerning the decay rate of $^{137}$Cs. In the Ellis experiment \citep{ell90}, a set of $^{137}$Cs standards was measured on a large, uncollimated, NaI system \cite{coh69} designed for low background, high spectral resolution counting of human subjects undergoing in-vivo neutron activation analysis (NAA). The counting system, described in detail in \citet{coh69}, was comprised of two planar, parallel 3$\times$9 arrays of 15 cm $\times$ 5 cm matched NaI(Tl) crystals with low-background phototubes, for a total of 54 detectors. This system was located in a heavily shielded, low-background room at BNL. Additional measures, such as a recirculating, filtered air system, were also incorporated to reduce background, as described by \cite{coh69}.

The $^{137}$Cs standard set referenced above, which was used to determine system stability and calibration, consisted of nine individual sources of 0.5 $\mu$Ci each. Over a six month period of initial set-up and calibration, as reported in \citet{coh69}, the standard deviation of the $^{137}$Cs counts was 0.54\%, implying good stability, and the background over the same period was 2.32$\times{}10^{4}\pm0.56\%$ counts/min over a $\sim$2.5 MeV spectrum window. The overall spectral resolution (ratio of peak width to spectrum-window width) for the $^{137}$Cs source geometry was 8.7\%. Careful controls to monitor and quickly correct for drift were maintained for the system to prevent counting errors. Additional corrections were calculated for geometry and other parameters which could lead to other systematic variations in the measured counting rate, as described in \citet{coh69}. The result was a very well designed, sensitive and stable counting system, with sensitivities of order 0.1 nCi. Calibration was performed on the system daily with the aforementioned $^{137}$Cs sources.

This counting system was utilized in conjunction with a broad beam neutron irradiator \cite{coh72}, for the in vivo NAA experiments described in \citet{ell90}, as mentioned above. The irradiator was specifically designed for high reproducibility \citep{coh73}, and comprised an array of fourteen 50 Ci encapsulated $^{238}$Pu/Be sources. In 1976, a second standard set was incorporated to be used on a weekly basis to monitor the reproducibility of the activation and counting systems together \citep{ell90}. This was a lucite rod which contained nine regions each of $\sim$10 cm$^3$ volume containing powdered manganese metal. The rod was placed in the neutron irradiator for 5 minutes, activated, and then  was placed in the counting bed and counted for fifteen minutes. When placed in the counting bed, the activated regions of the rod were in the same locations as the $^{137}$Cs sources with respect to the detector arrays \citep{ell90}.

Interestingly, the data collected on the activation product, $^{56}$Mn (T$_{1/2}$=2.5789 h, 100\% $\beta^{-}$-decay, \citep{Mn56}), showed fluctuations with an annual period, as reported by \citet{ell90}, and later confirmed by our group in \citet{jav10}. The important result from Ellis' work was that even though annual fluctuations appeared in the $^{56}$Mn data, they were not observed in the $^{137}$Cs standard, which was measured on the same detector system daily to confirm the calibration and efficiency of the detector system. It should be noted that both the Mn and Cs standards were contained in similar lucite fixtures. Therefore the thermal expansion of the lucite, which has an expansion coefficient of $\sim$7.1$\times$10$^{-5}$ cm/cm/$^{\circ}$C \citep{luc}, would have been the same for the two standards, and hence expansion due to temperature would not be a likely source of any seasonal variations. Furthermore, as described in \citet{jav10}, if the phase of the outdoor temperature is assumed to be similar to any annual variation in indoor temperature, it does not match the phase of the annual variation in the $^{56}$Mn data reported by \citet{ell90}.

Based on the care and thorough effort in the design, construction and calibration of the counting system in the Ellis experiment, aimed at reducing possible background and systematic variations, it seems probable that the observed differences in the fluctuations in the measured decay rates are intrinsic to the $^{56}$Mn and $^{137}$Cs decays themselves. Of course, the possibility also exists that something was changed in the activation process, either in the production of neutrons in the PuBe sources, or the (n,$\gamma$) capture process in the conversion of $^{55}$Mn to $^{56}$Mn. Therefore, in light of all of the examples previously mentioned where decay rate variations were observed in long-lived isotopes, the possibility that the $^{56}$Mn decay rate changed over the course of the year cannot be excluded.

In the spirit of the above analysis it is instructive to examine the PTB counting system where variations in the $^{133}$Ba count rate were observed, but not in $^{137}$Cs as described in Sec. \ref{sec:Analysis}. As previously noted, the isotopes were measured on a 4$\pi$ ionization chamber system, and the current generated in the ion chamber was measured with a Kiethley electrometer, model 6517A. This ionization chamber current is the result of the deposition of energy by the interaction of the decay photons of the isotope being measured in the detector, generally in the detector walls. Much of that energy is transferred to the working gas of the detector (in this case argon) via Compton electrons or photo-electrons. Direct interaction of photons in the gas itself can also occur, but even at a pressure of 2 MPa (20 atm), this probability is very low. However, this is not true for the lower energy Compton photons resulting from the original wall interactions.

One of the significant factors leading to the selection of ionization chambers in these types of measurements is their inherent stability with respect to systematic and environmental effects. It is well known that the output of the detector is relatively insensitive to changes in the detector bias voltage, since the voltage is high enough to prevent recombination of the electron/ion pairs generated, but too low to cause electron multiplication. Hence there is a fairly large width to this ``plateau'' such that the response is insensitive to voltage variations. Furthermore, as described in detail in \citet{jen10}, the construction of the detector is such that the geometry and functional parameters of the detector will experience only negligible changes as a result of changes to ambient variables such as temperature, pressure and humidity. Possible backgrounds and their seasonal variations were also analyzed in \citet{jen10} with respect to the annual fluctuations in the $^{226}$Ra measurements reported in \citet{sie98}. These were found to be too small an effect to account for the observed fluctuations in the $^{226}$Ra currents. It is reasonable to conclude that none of the known possible influences on the detector system were the likely cause of the fluctuations.

It should be pointed out that the previously discussed results of the analysis of $^{226}$Ra data acquired at PTB over the interval June 1999 to November 2008 differ from the earlier results of the analyses \citep{jen09b,stu10b} of data acquired at PTB over the interval November 1983 to October 1998 \citep{sie98}. Notably, the decay rate of $^{226}$Ra shows strong evidence of an annual oscillation in the earlier dataset but weaker evidence for such an oscillation in the later dataset (Figure \ref{fig:fig4}). In fact, \citet{sch10} shows that strong annual periodicities were also present in $^{85}$Kr, $^{108m}$Ag, $^{152}$Eu and $^{154}$Eu measured from 1990 to 1996. 

The difference between the two sets of measurements, as described by \citet{sch10}, is that the current measuring system used in the PTB measurements prior to October 1998 was a Townsend balance, but for data taken after 1998, a Keithley electrometer was used, as described in \citet{sch07,sch10}. With the introduction of the electrometer, the observed annual periodicity was reduced significantly. However, \citet{sch07} points out in his 2007 article (8 years after the switch to the electrometer) that ``for high accuracy measurements in metrology, a Townsend induction balance, i.e. a capacitor with voltage compensation, is used...''. While it remains to be seen which current measurement method will prove to be the more accurate, it is still evident--as noted in Sec. \ref{sec:Analysis}-- that there clearly remains an annual periodicity in the $^{133}$Ba data that is similar to those observed prior to the introduction of the electrometer, in addition to a weak one in the $^{226}$Ra data, as shown in Fig. \ref{fig:fig4}.

Further support for the existence of the annually-varying periodic behavior in the $^{133}$Ra data is presented in Table \ref{tbl:tbl1}, which lists other experiments where annual and sub-annual periodicities have been observed. What is important to note here is that 17 of the 23 results listed in Table \ref{tbl:tbl1} were collected by counting methods utilizing pulse processing from a variety of detector classes, and not the current measurement from an ion chamber. The systematics of pulse processing are significantly different from the measurement of small currents, such as those of the PTB ionization chamber, and are not subject to the same set of challenges and uncertainties.

\begin{table*}[ht]
\caption{Experiments where time-dependent decay rates have been observed.\label{tbl:tbl1}}
\small
\begin{center}
\begin{tabular}{ c c c c l c }
\hline
Isotope  & Decay Type &  Detector Type & Radiation Measured & Effect Observed & Reference \\
\hline
$^{3}$H & $\beta^{-}$ & Photodiodes & $\beta^{-}$ &  Periodicity: 1 yr$^{-1}$ & \cite{fal01}  \\
$^{3}$H & $\beta^{-}$ & Liquid Scintillator & $\beta^{-}$ &  Periodicity: 1/d, 12.1 yr$^{-1}$, 1 yr$^{-1}$ & \cite{shn98}  \\
$^{3}$H & $\beta^{-}$ & Liquid Scintillator & $\beta^{-}$ &  Periodicity: $\sim$12.5 yr$^{-1}$ & \cite{vep12}  \\
$^{3}$H & $\beta^{-}$ & Solid State (Si) &$\beta^{-}$ &   Periodicity: $\sim$2 yr$^{-1}$ & \cite{lob99} \\
$^{22}$Na/$^{44}$Ti$^{\left[a\right]}$ & $\beta^{+},\kappa$ & Solid State (Ge) & $\gamma$ & Periodicity: 1 yr$^{-1}$ & \cite{oke12} \\
$^{36}$Cl & $\beta^{-}$ & Proportional &$\beta^{-}$ &   Periodicity: 1 yr$^{-1}$, 11.7  yr$^{-1}$, 2.1 yr$^{-1}$ & \cite{jen09b,stu10a,stu11a} \\
$^{36}$Cl & $\beta^{-}$ & Geiger-M\"{u}ller  &$\beta^{-}$ &   Periodicity: 1 yr$^{-1}$ & \cite{jen12} \\
$^{54}$Mn & $\kappa$ &  Scintillation  & $\gamma$ &   Short term decrease during solar flare & \cite{jen09a} \\
$^{54}$Mn & $\kappa$ &  Scintillation  & $\gamma$&   Periodicity: 1 yr$^{-1}$ & \citep{jen11}\\
$^{56}$Mn & $\beta^{-}$ &  Scintillation &$\gamma$ &   Periodicity: 1 yr$^{-1}$ & \cite{ell90} \\
$^{60}$Co & $\beta^{-}$ & Geiger-M\"{u}ller  &$\beta^{-}$,$\gamma$ &   Periodicity: 1 yr$^{-1}$ & \cite{par10a,par10b} \\
$^{60}$Co & $\beta^{-}$ & Scintillation &$\gamma$ &   Periodicity: 1/d, 12.1 yr$^{-1}$ & \cite{bau07} \\
$^{85}$Kr & $\beta^{-}$ & Ion Chamber &$\gamma$  & Periodicity: 1 yr$^{-1}$ & \cite{sch10} \\
$^{90}$Sr/$^{90}$Y & $\beta^{-}$ & Geiger-M\"{u}ller  &$\beta^{-}$ &  Periodicity: 1 yr$^{-1}$, 11.7  yr$^{-1}$ & \cite{par10a,par10b,stu12a} \\
$^{108m}$Ag & $\kappa$ & Ion Chamber &$\gamma$  & Periodicity: 1 yr$^{-1}$ & \cite{sch10} \\
$^{133}$Ba & $\beta^{-}$ & Ion Chamber & $\gamma$ &   Periodicity: 1 yr$^{-1}$ & This work \\
$^{137}$Cs & $\beta^{-}$ & Scintillation & $\gamma$ &   Periodicity: 1 d$^{-1}$, ~12.1 yr$^{-1}$ & \cite{bau07}\\
$^{152}$Eu & $\beta^{-},\kappa$  & Solid State (Ge) & $\gamma^{\left[b\right]}$ & Periodicity: 1 yr$^{-1}$ & \cite{sie98} \\
$^{152}$Eu & $\beta^{-},\kappa$ & Ion Chamber &$\gamma$ & Periodicity: 1 yr$^{-1}$ & \cite{sch10} \\
$^{154}$Eu & $\beta^{-},\kappa$ & Ion Chamber &$\gamma$ & Periodicity: 1 yr$^{-1}$ & \cite{sch10} \\
$^{222}$Rn$^{\left[c\right]}$ & $\alpha,\beta^{-}$  & Scintillation &$\gamma$ & Periodicity: 1 yr$^{-1}$, 11.7 yr$^{-1}$, 2.1 yr$^{-1}$ & \cite{ste11,stu12b} \\
$^{226}$Ra$^{\left[c\right]}$ & $\alpha,\beta^{-}$ & Ion Chamber &$\gamma$ & Periodicity: 1 yr$^{-1}$, 11.7 yr$^{-1}$, 2.1 yr$^{-1}$ & \cite{jen09b,stu10b,stu11a} \\
$^{239}$Pu & $\beta^{-}$ & Solid State  &$\alpha$ &   Periodicity: 1/d, 13.5 yr$^{-1}$, 1 yr$^{-1}$ & \cite{shn98} \\
\hline
\multicolumn{6}{l} {$^{\left[a\right]}$\footnotesize{}Only the count rate ratio data were available.} \\
\multicolumn{6}{l} {$^{\left[b\right]}$\footnotesize{}Only the $\kappa$ photon was measured.} \\
\multicolumn{6}{l} {$^{\left[c\right]}$ \footnotesize{}Decay chain includes several primarily $\beta$-decaying daughters which also emit photons.} \\
\hline
\end{tabular}
\end{center}
\end{table*}

An additional important feature of the data in Table \ref{tbl:tbl1} is that there are ten in the list that exhibit sub-annual periodicities in addition to the annual variations. While it is possible to attribute the annual periods to a ``seasonal'' influence with a clear annual variation, such as temperature, periodicities on the order of six months, one month, or less, are not. Furthermore, as shown in Refs. \citep{stu10a,stu10b,stu11a,shn98,vep12}, the observed decay frequencies align closely with known solar periodicites. This tends to support the original hypotheses of \citet{jen09a}, \citet{jen09b}, and \citet{fis09}, suggesting there is a solar influence on radioactive decay rates on Earth. Even more interesting are the results presented by our group in \citet{stu12a} which indicate that the data originally presented in \citet{par10a}, and \citet{par10b}, exhibit a frequency structure similar to solar diameter measurements from the Mt. Wilson Solar Observatory. Thus, the evidence supporting a putative solar evidence seems at least as reasonable as simply attributing the decay rate variations to unspecified environmental or systematic effects.

Further support for concluding that the source of the variations being observed is not simply an environmental or systematic effect can be found in the $^{137}$Cs and $^{133}$Ba data presented in Sec. \ref{sec:Analysis}. These measurements were made in the same time frame, with the same detector, and with the same current measuring electronics. It is difficult to produce a viable conventional scenario to account for this difference, particularly in light of the results of \citet{ell90}. Similar results were observed in an experiment by  \citet{alb86}, who measured $^{32}$Si (T$_{1/2}$=153 y, 100\% $\beta^{-}$-decay \citep{Si32}) and $^{36}$Cl (T$_{1/2}$=301,000 y, 98.10\% $\beta^{-}$-decay, 1.90\% K-capture \citep{Cl36}) alternately on a differential gas proportional detector system \citep{har73} using 30 minute counts each for ten hours total. For each isotope, the 30-minute counts were aggregated into an integral count for the day. By taking the ratio of the two integrated counts, i.e., $^{32}$Si/$^{36}$Cl, systematic and environmental effects should have been canceled out. However, the ratio exhibited an annual periodicity, which matched, both in amplitude and phase (for the period the two experiments overlapped), the $^{226}$Ra data taken at PTB \citep{sie98}, as reported in \citet{jen09b}. Of particular importance in the BNL data is that even though the two isotopes, $^{32}$Si and $^{36}$Cl, were measured on the same detector on the same days, the decay data exhibited different structure, in both the amplitude and phase of the respective measured decay data, as discussed by our group in \citet{jav10} and \citet{stu11b}.

\section{Conclusions}

Our purpose here has been to present new data, for $^{133}$Ba and $^{137}$Cs, which were measured on the same detector system for the same time period, where one isotope ($^{133}$Ba) exhibited a clear annual periodicity and the other ($^{137}$Cs) did not. This result, in addition to the results of decay experiments listed in Table \ref{tbl:tbl1}, indicates that the failure to observe the annual (or other) periodicity in one isotope does not exclude that possibility in others. In light of Table \ref{tbl:tbl1}, we can state in general that our studies to date suggest the following: (a) not all nuclides exhibit variability in decay constants;  (b) among nuclides that do exhibit this variability, the patterns of variability (e.g., amplitude and phase of any oscillation) are not all the same; and (c) for nuclides that do exhibit variability, the patterns themselves may vary over time. More specifically, the results presented in Table \ref{tbl:tbl1} in support of time-varying nuclear decay rates cumulatively represent over sixty years of data collection, in comparison to the 0.5 years of data from \citep{bel12}.

Clearly, a new series of experiments with a variety of nuclides and a variety of detectors would help to determine definitively whether the amplitudes and phases of annual oscillations are steady or variable and, if variable, the characteristics of the variability. This could lead to a determination of the cause of the non-random behavior exhibited by the decays of some isotopes, which should eventually lead to an understanding of the governing mechanism.

\section*{Acknowledgements}
We are indebted to the staff at Physikalisch-Technische Bundesanstalt for generously making their data available to us, and we owe special thanks to Dr. Heinrich Schrader for his support and advice.





\bibliographystyle{model1a-num-names}







\end{document}